\documentclass[sigconf]{aamas}    

\usepackage{balance} 
\usepackage{graphicx}
\usepackage{multirow}
\usepackage{subcaption}
\usepackage{xcolor}
\usepackage{xspace}

\setcopyright{ifaamas}
\acmConference[AAMAS '26]{Proc.\@ of the 25th International Conference
on Autonomous Agents and Multiagent Systems (AAMAS 2026)}{May 25 -- 29, 2026}
{Paphos, Cyprus}{C.~Amato, L.~Dennis, V.~Mascardi, J.~Thangarajah (eds.)}
\copyrightyear{2026}
\acmYear{2026}
\acmDOI{}
\acmPrice{}
\acmISBN{}

\title[AAMAS-2026 Formatting Instructions]{Asynchronous Multi-Agent Reinforcement Learning for 5G Routing under Side Constraints}

\author{Sebastian Racedo and Brigitte Jaumard}
\affiliation{
  \institution{Concordia University}
  \city{Montreal (Qc)}
  \country{Canada}}
\email{brigitte.jaumard@concordia.ca}

\author{Oscar Delgado}
\affiliation{
  \institution{\textit{Ecole de Technologie Supérieure (ETS)}}
  \city{Montreal (Qc)}
  \country{Canada}}

\author{Meysam Masoudi}
\affiliation{
  \institution{Ericsson}
  \city{Kista}
  \country{Sweden}}

\begin{abstract}
Networks in the current 5G and beyond systems increasingly carry heterogeneous traffic with diverse quality-of-service constraints, making real-time routing decisions both complex and time-critical. 
A common approach, such as a heuristic with human intervention or training a single centralized RL policy or synchronizing updates across multiple learners, struggles with scalability and straggler effects. 
We address this by proposing an asynchronous multi-agent reinforcement learning (\NameProposal) framework in which independent PPO agents, one per service, plan routes in parallel and commit resource deltas to a shared global resource environment. 
This coordination by state preserves feasibility across services and enables specialization for service-specific objectives. 
We evaluate the method on an O–RAN like network simulation using nearly real-time traffic data from the city of Montreal. 
We compared against a single-agent PPO baseline. 
{\NameProposal} achieves a similar Grade of Service (acceptance rate) (GoS) and end-to-end latency, with reduced training wall-clock time and improved robustness to demand shifts. 
These results suggest that asynchronous, service-specialized agents provide a scalable and practical approach to distributed routing, with applicability extending beyond the O-RAN domain.
\end{abstract}


\newcommand{\BibTeX}{\rm B\kern-.05em{\sc i\kern-.025em b}\kern-.08em\TeX}

\def \NameProposal  {AMARL} 
\def \RLBaseline    {SARL}

\begin{document}


\pagestyle{fancy}
\fancyhead{}


\maketitle 


\section{Introduction}
\label{sec:introduction}

Modern programmable networks---encompassing 5G and beyond---must concurrently serve heterogeneous services with markedly different quality-of-service (QoS) requirements, such as high throughput Enhanced Mobile Broadband (eMBB) and Ultra-Reliable Low Latency Communication (URLLC). 
Meeting these diverse constraints over shared infrastructures turns routing into a moving target: 
traffic, contention, and link conditions evolve quickly and non-uniformly across space and time. 
Static routing and classic traffic engineering approaches struggle to deliver per-service QoS under nonstationary demand and partial observability, motivating learning-based control that can adapt online to congestion and service priorities \cite{Chunlei_2020, Almasan_2022}. 
Furthermore, given the rapid traffic shifts and partial observability, a non-learning baseline (e.g., a fixed shortest-path or periodically re-optimized routing scheme) would lack the adaptivity to handle changing conditions in real time. Prior work has shown that purely heuristic or planning-based methods often require heavy recomputation and still cannot react quickly to unforeseen changes \cite{Yu_2023}. 
In contrast, a trained Reinforcement Learning (RL) agent can adapt on the fly with negligible decision overhead once deployed. For this reason, we focus our investigations on learning-based controllers.

A centralized reinforcement-learning controller is a natural first step \cite{Boyan_1993}, but quickly runs into fundamental limits:
the global state and action space grows combinatorially with network size; decisions for different services are tightly coupled; and any single learner becomes a synchronization bottleneck under high arrival rates \cite{Oroojlooy_2022}. Moreover, real deployments make decisions at different granularities and timescales (per-flow, per-burst, per-service), so strict coordination is often impractical.

Asynchrony aligns with real deployments (agents observe and act at different cadences), avoids straggler bottlenecks (e.g., performance bottleneck in distributed computing where the overall speed of a job is dictated by the slowest task), and improves sample throughput by decoupling rollout from learning \cite{Mnih_2016, Espeholt_2018}. Specializing agents by service lets each policy target its QoS envelope (e.g., tight URLLC latency vs. eMBB throughput) while still cooperating through the shared state.

In the context of networking, and more specifically, Open RAN (O-RAN), networks that follow this architecture enable near-real-time (near-RT) closed-loop control through the RAN Intelligent Controller (RIC) and third-party xApps operating over open interfaces (A1, E2). 
Near-RT xApps can observe service demand, link conditions, and QoS metrics on 10–100\, ms timescales and issue control actions (e.g., routing/steering) accordingly; Non-RT rApps provide slower-timescale policy and model management \cite{Bonati_2021, Elyasi_2025}. 
This architecture naturally supports distributed agents specialized by service or domain, coordinated indirectly via shared state and policies---and thus is a compelling domain for asynchronous Multi Agent RL (MARL)s. While there is a growing body of RL for O-RAN (e.g., slicing, scheduling), RL routing at the RAN/transport layer is comparatively nascent.

Despite rapid progress in MARL and O-RAN automation, 
there is little work on fully asynchronous, service-specialized MARL for routing in a shared environment, where independent agents, one per service, learn in parallel while coupling only via shared network state. 
Most prior efforts either 
\textit{(i)}    centralize training/execution, 
\textit{(ii)}   synchronize agent updates/actions, or 
\textit{(iii)}  address other RAN control problems rather than routing. 
We close this gap by proposing an asynchronous multi-agent RL "{\NameProposal}," framework where independent Proximal Policy Optimization (PPO) \cite{Schulman_2017} agents plan and commit per-request route decisions to a shared environment that enforces capacity and latency constraints. 
In this vein, asynchronous coordination means each agent observes and acts on its own cadence without waiting for others, in contrast to synchronized steps. This design mirrors real deployments where agents may have different reaction times or processing delays. In a synchronous scheme, faster agents must idle while waiting for slower ones, which creates straggler bottlenecks and underutilized resources. The O-RAN setting serves as both motivation and a realistic case study. We evaluate our approach using a realistic 24-hour traffic simulation for the city of Montreal. To isolate and quantify the benefits of asynchronous multi-agent vs a single agent, we compare its performance with "{\RLBaseline}", a general single routing policy. Our contributions are as follows:
\begin{itemize}
    \item Asynchronous, service-specialized MARL for routing. We formulate service routing in a shared network as a MARL problem with one agent per service, and design an asynchronous training/execution pipeline that avoids global barriers while preserving feasibility through a shared state.
    \item Coordination-by-state with action masking. Agents specialize to service QoS and coordinate implicitly via resource contention in the shared environment.
    \item O-RAN case study and evaluation. Using an O-RAN like routing simulation network, we compare against a strong centralized single-agent baseline and show consistent performances in terms of Grade of Service (acceptance) and end-to-end latency, alongside lower training wall-clock time.
\end{itemize}

The remainder of this paper is organized as follows. 
Section \ref{sec:literature_review} briefly reviews past work on RL, MARL, and Asynchronous RL.
Section \ref{sec:problem_formulation} formalizes the shared-environment, service-specific routing problem formulation.
Section \ref{sec:methodology} presents the asynchronous multi-agent PPO design and implementation. 
Section \ref{sec:setup} details the O-RAN simulation and traffic generator. 
Section \ref{sec:results} presents the performance evaluation and discussion of the results, along with their implications.  
Section \ref{sec:conclusions_FW} concludes the paper and outlines future work.


\section{Literature Review}
\label{sec:literature_review}

\subsection{General MARL}
 Centralized single-agent controllers face scalability and observability limits as network size and state–action spaces grow; a single policy must reconcile conflicting objectives and is prone to straggler effects when decisions serialize. MARL decentralizes decision making across agents acting in parallel, but introduces non-stationarity (each agent’s environment shifts as others learn) and coordination challenges \cite{Busoniu_2008, Oroojlooy_2022}. Nonetheless, recent results have shown that even simple on-policy methods can perform strongly in cooperative MARL. In particular, a multi-agent variant of PPO (sometimes called MAPPO) achieved surprisingly competitive performance and sample efficiency compared to more complex off-policy algorithms \cite{Yu_2022}. In fact, with minimal tuning or specialized architectures, PPO-based agents matched or outperformed state-of-the-art methods, suggesting that PPO can serve as a strong baseline in cooperative multi-agent tasks.


\subsection{Asynchronicity}

Very few work as been done combining RL and asynchronicity. Asynchronous training mitigates stragglers and improves sample throughput by decoupling rollout from learning.
In single-/few-agent deep RL, \emph{Asynchronous Advantage Actor–Critic} (A3C) established that parallel actor–learners stabilize training and deliver near-linear wall-time speedups \cite{Mnih_2016}. 
IMPALA further decouples acting and learning by utilizing centralized learners and distributed actors, employing V-trace for stable off-policy correction at high throughput \cite{Espeholt_2018}. 
Distributed prioritized replay (Ape-X) also demonstrates that many asynchronous actors can accelerate training without hurting final performance \cite{Horgan_2018}. Beyond speeding up training, asynchronous coordination can yield tangible benefits in real-world multi-agent tasks. For example, in multi-robot exploration, a fully synchronous MARL formulation (where all agents must move in lock-step) was found to be time-inefficient due to different agents’ action durations \cite{Yu_2023}. Yu \textit{et al.} \cite{Yu_2023} proposed an asynchronous MARL approach (ACE) that lets each robot act as soon as it is ready eliminated this bottleneck – significantly reducing total exploration time compared to both a classical planning method and a synchronous RL baseline. 
These systems motivate our {\NameProposal} setting, where independent service agents learn concurrently and implicitly coordinate through a shared environment, rather than through synchronized multi-agent updates.

\subsection{RL in networking}

In networking, RL has been explored for QoS-aware path selection, experience-driven traffic engineering, and hop-by-hop forwarding in wired and wireless networks. Early work embedded reinforcement learning directly into routers (\emph{Q-routing}), showing that local, experience-driven forwarding can reduce end-to-end delay relative to shortest-path routing \cite{Boyan_1993}.
With deep function approximation, RL has been applied to routing and traffic engineering in centralized (SDN) and distributed settings, learning to react to congestion and topology dynamics \cite{Chunlei_2020, Almasan_2022, Jaumard_2024_GNNet}.
SDN controllers with Deep RL (DRL) improved delay/utility by learning TE policies end-to-end, rather than solving per-epoch optimizations \cite{Xu_2018}. 
Recent work augments RL with graph encoders and causal attributions to generalize across topologies and improve credit assignment \cite{He_2024, Almasan_2022}. 
Decentralized MARL for routing (e.g., mobile ad hoc and overlay networks) shows gains in packet delivery and latency by training per-node policies that leverage local observations and limited messaging, while highlighting stability and safety constraints as open issues \cite{Boyan_1993, Chunlei_2020}.

Compared to prior networking work that is centralized or synchronous, our {\NameProposal} brings asynchrony to service specific routing in an O-RAN use case setting. Concretely: (i) independent PPO agents---one per service---route flows in parallel on local episodic copies and commit bandwidth/compute deltas to a shared environment; (ii) we show that this design delivers substantial wall clock gains (about 30\% faster training and 15\% faster evaluation) without loss of QoS relative to a baseline; and (iii) it confers operational advantages absent in monolithic policies such as fault isolation across services, and modular specialization/upgrades.



\section{Problem Formulation}
\label{sec:problem_formulation}

At its core, the routing problem can be viewed as a large-scale, multi-commodity network optimization: deciding which flows to admit and how to route them to maximize service acceptance while meeting latency and capacity constraints. Solving this optimally (e.g., via Mixed-Integer Linear Programming (MILP)) is NP-hard and impractical for real-time control, especially as network size grows and conditions change. Instead, we embed this optimization into a MARL framework – formulating per-service routing as a sequential decision problem. Each agent learns to make routing decisions that collectively satisfy the global constraints, essentially optimizing the network’s resource allocation on the fly. Following is a formal definition of the problem.

Let the network be a directed graph \(G=( V, E)\) with nodes \( V= V_c\cup\ V_n\) (compute-capable nodes and switches) and directed links \(E\). Each physical bidirectional link is modeled as two directed edges (UpLink (UL) and a DownLink (DL) connections); for \(e\in E\), let \(C_e>0\) be capacity (Mbps) and \(t_e\ge 0\) propagation delay (ms). Each compute node \(v\in V_c\) has usable compute budget \(\kappa_v>0\) (e.g., CPU units); we treat \(\kappa_v\) as the fraction of physical capacity allowed for admission (e.g., \(80\%\)).

Let \(S\) be the set of service classes. Each service \(s\in S\) is associated with an ordered service function chain (SFC)
\[
\boldsymbol f_s=(f_{s,1},\ldots,f_{s,n_s}),\qquad f_{s,k}\in F,\quad K_s=\{1,\dots,n_s\},
\]
where \(\ F=\{\text{O-RU},\text{O-DU},\text{O-CU},\text{UPF}_x\}\) is the catalog of VNFs and \(n_s\) is the chain length for service \(s\). We also use the segment index set \(\bar{K}_s=\{0,\dots,n_s\}\) to denote path segments between consecutive waypoints (source\(\to f_{s,1}\), \(f_{s,1}\to f_{s,2}\), \(f_{s,3}\to\)destination). The placement is fixed. Let \(\bar x_{v,f}\in\{0,1\}\) indicate whether function \(f\) is instantiated/hosted at node \(v\). This initial placement is precomputed and considered a parameter of the problem (not a decision).

Let \(P\) be the set of flow requests. Each \(p\in P\) belongs to a service \(s(p)\in S\) and is characterized by source \(s_p\in V\), destination \(d_p\in V\), bandwidth \(b_p>0\), and end-to-end latency budget \(\tau_p>0\). The traffic of \(p\) must traverse \(\boldsymbol f_{s(p)}\) in order.

We use the following decision variables:
\begin{itemize}
    \item \(\text{acc}_p\in\{0,1\}\): admission of request \(p\).
    \item \(a_{v,k,p}\in\{0,1\}\): request \(p\) executes SFC stage \(k\) of service \(s(p)\) at node \(v\) (i.e., \(v\) hosts \(f_{s(p),k}\)).
    \item \(y_{e,p,k}\in\{0,1\}\): link \(e\) is used by request \(p\) along segment \(k\!\in\!\bar{ K}_{s(p)}\) (from stage \(k\) to \(k{+}1\), with \(k{=}0\) for \(s_p\!\to\!f_{s(p),1}\) and \(k{=}n_{s(p)}\) for \(f_{s(p),n_{s(p)}}\!\to\!d_p\)).
\end{itemize}


We used the following constraints:
\begin{itemize}
    \item Each stage is executed exactly once and only at a node where the function is placed:
        \begin{alignat}{2}
            & \sum_{v\in V_c} a_{v,k,p} \;=\; \text{acc}_p, \qquad 
            && p\in P,\ k\in K_{s(p)}, 
            \label{eq:sfc_assign_one}
            \\
            & a_{v,k,p} \;\le\; \bar x_{v, f_{s(p),k}},
            &&  k\in K_{s(p)}, p\in P, \ v\in V_c.
            \label{eq:sfc_hosting}
        \end{alignat}

    \item For every link $e\in E$, the sum of admitted segment flows cannot exceed capacity:
    \begin{align}
        \sum_{p\in\ P}\ \sum_{k\in\bar{K}_{s(p)}} b_p\, y_{e,p,k} \;\le\; C_e,
        \qquad e\in E.
        \label{eq:link_capacity}
    \end{align}

    \item For each compute node $v\in\mathcal{V}_c$, the total compute consumed by admitted requests assigned to $v$ cannot exceed its set capacity (it can not be above 80\% utilization for the CPU, this is a standard practice in Data Centers (DCs)):
    \begin{align}
        \sum_{p\in P}\ \sum_{k\in K_{s(p)}} b_p\, c_{v, f_{s(p),k}}\, a_{v,k,p}
        \;\le\; \kappa_v,
        \qquad v\in V_c,
        \label{eq:compute_capacity}
    \end{align}
    where \(c_{v,f}\) is the per-Mbps compute demand of \(f\) at \(v\).

    
\end{itemize}
The overall objective is to learn a set of polices $\pi$ (one for each service) that are optimal with respect to our defined reward function. The goal is not to find all possible valid paths, but to identify the one that maximizes the cumulative reward, which encompasses balancing acceptance, latency, and violations.


\section{Asynchronous Multi-Agent Design}
\label{sec:methodology}
We design an asynchronous MARL system in which $S$ independent agents $\{\mathcal{A}_s\}_{s=1}^{S}$ (one per service, here $S{=}6$) interact in parallel with a single shared environment $\mathcal{E}^\star$ that maintains the global network state (link utilization, compute utilization, time, and per-request logs). Each agent handles only the flows of its service type, runs one local episodic copy $\widetilde{\mathcal{E}}_s$ per flow, and upon success emits a $\Delta(p)$ that is applied to $\mathcal{E}^\star$. Agents never synchronize with each other; coordination arises through resource contention in $\mathcal{E}^\star$.


\subsection{Agents, Environment, and Asynchrony}
\begin{figure}[ht]
        \centering
        \includegraphics[width=1.0\linewidth]{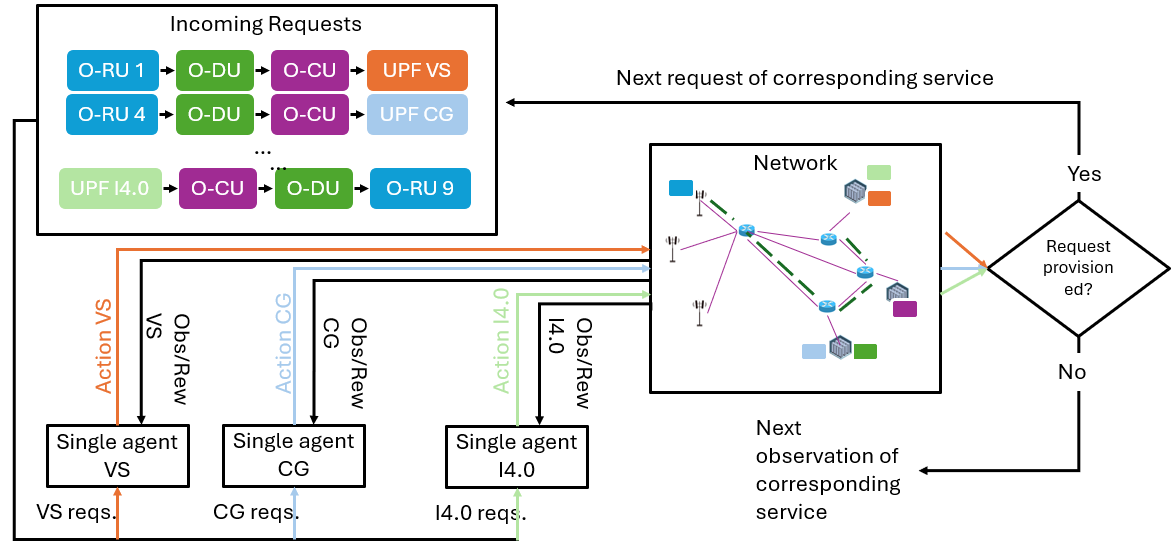}
        \caption{Asynchronous Multi-Agent Diagram}
        \label{fig:AMARL_Diagram}
        \Description{AMARL agent diagram}
    \end{figure}

    Figures \ref{fig:SARL_Diagram} and \ref{fig:AMARL_Diagram} compare the centralized single-agent pipeline (used as a baseline, referred to as \RLBaseline) with our asynchronous multi-agent design. Concretely:

    \begin{figure}[ht]
        \centering
        \includegraphics[width=0.7\linewidth]{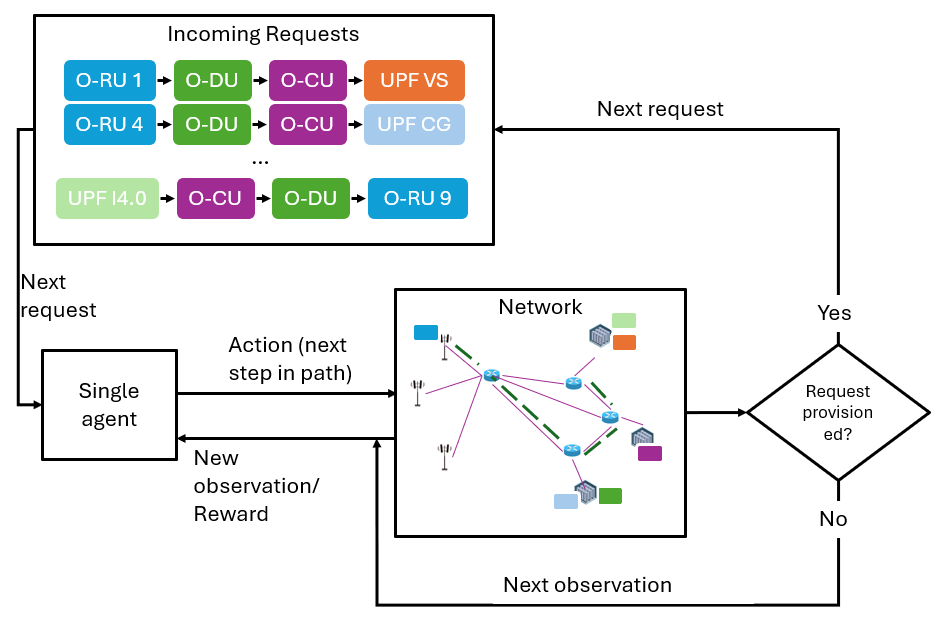}
        \caption{Single Agent Diagram}
        \label{fig:SARL_Diagram}
        \Description{Single agent diagram}
    \end{figure}

    From Figure \ref{fig:SARL_Diagram}, it can be seen that a single agent learns a non-specialized policy for all the flows that are coming into the network, regardless of which service they correspond to. On the other hand, Figure \ref{fig:AMARL_Diagram} shows that there are a given number of agents, where each one learns its own specialized policy based on the flows corresponding to that service. Each agent $\mathcal{A}_s$ executes the following loop independently:

    \begin{enumerate}
        \item Fetch next request $p\!\sim\!{D}_s$ for service $s$.
        \item Instantiate local env $\widetilde{{E}}_s$ with a \emph{snapshot} of the shared state
        $\mathsf{state}({E}^\star)$ (current $u_e$, $u_v$, time index, and per-service flows).
        \item one local episode to provision the SFC of $p$ step-by-step.
        \item Emit $\Delta(p)$ if successful:
        \[
        \Delta(p)=\big\{\{\Delta u_e\}_{e\in {E}},\ \{\Delta u_v\}_{v\in {V}_c},\ \mathsf{log}(p)\big\},
        \]
        where $\Delta u_e{=}b_p$ on links used by $p$ and $\Delta u_v{=}b_p\, c_{v,f}$ on VNF execution nodes. Apply $\Delta(p)$ to ${E}^\star$ via a lock-guarded commit(): if capacity checks would violate \eqref{eq:link_capacity} or \eqref{eq:compute_capacity} after the commit, commit() aborts and the episode is marked rejected.
    \end{enumerate}

    The propagation delay for $p$ is accumulated in the local episode as the number of links used (each link has a latency associated with it), and the processing delay is the accumulated processing time needed on each VNF to run the necessary functions given the bandwidth of each flow   
    using the $u_v$ from the snapshot. A commit succeeds only if the realized end-to-end latency is less than the flow max latency budget; otherwise the episode terminates with rejection. Upon a successful commit, $\mathcal{E}^\star$ updates $u_e,u_v$ and logs the achieved latency for evaluation.

    We implement ${A}_s$ as Ray \texttt{Task}s (PPO Trainers) with Rllib \cite{Liang_2021} and ${E}^\star$ as a Ray \texttt{Actor} using Ray Core \cite{Moritz_2018}.
    
    \subsection{Observations, Actions and Mask}

    At step $t$ while routing request $p$, the agent observes a dictionary $o_t=\{\texttt{obs},\ \texttt{node\_feat}\}$:

    Obs vector ($\texttt{obs}$):
    \begin{itemize}
        \item Bandwidth $b_p$, remaining budget $\hat{\tau}_t$ (initially $\tau_p$ minus accumulated delay).
        \item Segment index $k_t\in\{0,\dots,K_p\}$; current node id $i_t$ and segment target type (e.g., $f_{p,k_t+1}$ or destination).
        \item Graph distance (e.g., shortest path hops) from $i_t$ to the nearest feasible node hosting $f_{p,k_t+1}$ (or to $d_p$ if $k_t{=}K_p$).
    \end{itemize}

    Node features ($\texttt{node\_feat}$):
    For each node $v$: (i) multi-hot role (can host O-DU/O-CU/UPF\_$x$), (ii) normalized degree, (iii) current free compute
    , and (iv) indicator if $v$ is a candidate for the next SFC stage. To mitigate partial observability without a global state dump, we include a summary, which we call \texttt{global} from the snapshot: link utilization, free compute per tier (O-DU/O-CU/UPF), and time-of-day index. These are scalars appended to the policy input.

\begin{figure}
        \centering
        \includegraphics[width=0.5\linewidth]{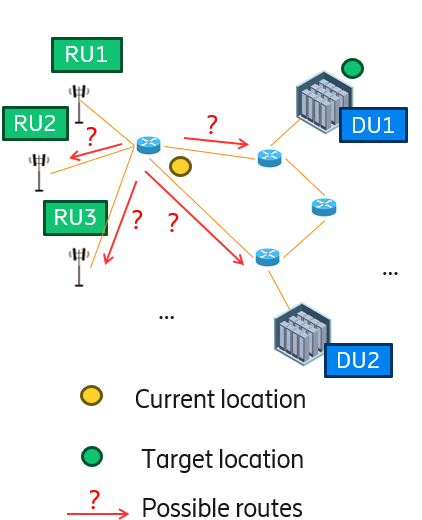}
        \caption{Example of step where the current segment is 0 (src is RU1 and target is DU1)}
        \label{fig:example_step}
        \Description{step example}
    \end{figure}

    The action space is defined as a discrete action $a_t\in\{0,\dots, D_{\max}\}$. At step $t$, the agent chooses the next hop from the current node $i_t$. Let ${N}(i_t)$ be the neighbor set and $D_{\max} \ge \max_i |{N}(i)|$. Figure \ref{fig:example_step} shows a high-level view of a step. 

    We expose a binary mask (`action\_mask`) $m_t \in \{0,1\}^{D_{\max}}$ indicating which actions are valid in the current state of an agent. The environment generates this mask at each step to mitigate each agent from selecting actions that would violate hard constraints.


    \subsection{Reward Structure}

    To guide the agent toward QoS satisfaction while maintaining efficient paths, the total reward for an episode is the sum of stepwise components plus a terminal bonus. Let $\tau$ denote the trajectory (one flow). The agent maximizes the discounted return, where the conceptual total reward is:
    \begin{equation}
    \label{eq:reward_routing_only}
    RW_{\text{total}}(\tau) \;=\; RW_{\text{terminal}}
    \;+\; \sum_{t \in \tau} \Big( rw_{\text{shaping}} \;+\; rw_{\text{intermediate}}\Big).
    \end{equation}
    Each term is defined as follows:

    \begin{itemize}
      \item \textbf{Shaping Reward ($rw_{\text{shaping}}$):} At every action, the agent receives a small reward proportional to the reduction in (estimated) remaining distance or latency to the current waypoint (next required function or final destination). A penalty is applied for cycles/backtracking to discourage loops.
    
      \item \textbf{Intermediate Success Reward ($rw_{\text{intermediate}}$):} When a segment is completed (the agent reaches any node hosting the required function for that segment, or the destination at the final segment), a positive bonus is given. This bonus is penalized by the number of hops spent within the segment to encourage path efficiency.
    
      \item \textbf{Terminal Reward ($RW_{\text{terminal}}$):} At episode end, a positive reward is granted if the request is successfully provisioned within its latency budget and capacity constraints; a significant negative penalty is given otherwise (e.g., no feasible next hop or budget exceeded). This makes acceptance the dominant learning signal.
    \end{itemize}

Each agent uses a service-specific policy with a Graph Convolutional Networks (GCNs) backbone over \texttt{node\_feat} (3 GCNConv layers), concatenated with the embedded \texttt{obs} and \texttt{global} vectors, followed by a shared MLP and two heads (actor logits, critic value) (more on the values used for training can be seen in Section \ref{sec:setup}). 


\section{Experimental Setup}
\label{sec:setup}


\subsection{Network Topology}

\begin{figure*}[ht]
        \centering
        \includegraphics[width=1.0\textwidth]{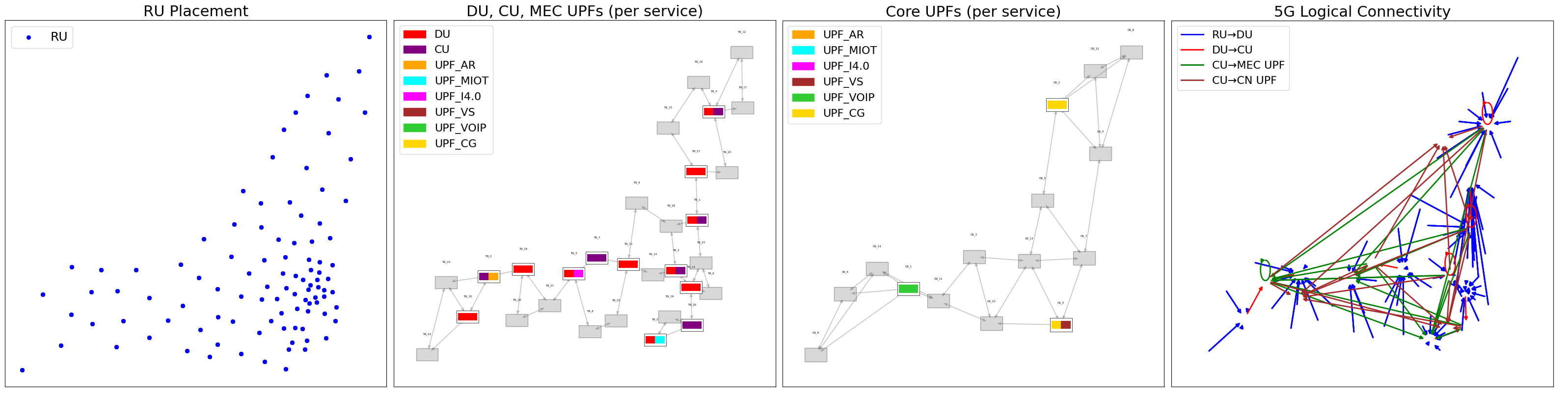}
        \caption{Fixed placement of logical functions capable of accommodating a dynamic traffic over 24 hours}
        \label{fig:placement_result_pcenter}
        \Description{Network figure}
    \end{figure*}

    Our scenario consists of 300 O-RUs, 10 O-DUs, 6 O-CUs, and 7 UPFs connected over the transport/core network across a 5G network. 
    Figure \ref{fig:placement_result_pcenter} shows the initial placement of logical functions used throughout all experiments. This placement was precomputed to be feasible for a full 24-hour demand profile. To make the routing problem meaningful, we guarantee that there exists more than one possible path for the majority of the flows (that satisfy the latency constraint); the learning agents then solve per-flow routing between the pre-placed SFC waypoints under the shared bandwidth/compute constraints.

    Tables \ref{tab:latency_stats_training} and \ref{tab:latency_stats_test} summarize, per service, the characteristics of the training and test traffic. The details of what each column means are as follows:
    \begin{itemize}
        \item Flows: The number of individual requests of that service in the split. 
        \item Mean Latency Budget (ms): is the target end-to-end deadline or maximum allowed latency budget configured for that service class (service-level objective).
        \item Mean Processing Delay (ms): is the average cumulative VNF processing time incurred per flow along its SFC (measured at the O-RU/O-DU/O-CU/UPF nodes actually traversed).
        \item Mean Margin (ms): is the average slack available for non-processing delay (average propagation delay budget), computed as
            \[
            \text{Mean Margin} = \text{Mean Latency Budget} - \text{Mean Processing Delay},
            \]
        \item Mean Total Bandwidth (Mbps): is the average aggregate offered bandwidth for that service over the 24\,h horizon (per time slot), and indicates how much capacity that service typically consumes network-wide.
    \end{itemize}
    
    \begin{table}[ht]
    \centering
    \caption{Detailed Latency and Processing Statistics Per Service (Training Data)}
    \label{tab:latency_stats_training}
    \resizebox{\columnwidth}{!}{
    \begin{tabular}{|c|r|c|c|c|c|}
        \hline 
        \multirow{2}{*}{\textbf{Services}} & \multirow{2}{*}{\textbf{Flows}} & \textbf{Mean Latency} & \textbf{Mean Proc.} & \textbf{Mean Margin} & \textbf{Mean Total} \\
         & & \textbf{Budget (ms)} & \textbf{Delay (ms)} & \textbf{(ms)} & \textbf{Bandwidth (Mbps)} \\
        \hline\hline
        Augmented reality   & \phantom{1}2,100  & \phantom{1}20.0  & 5.21 & 14.79              & 1,210.64 \\
        Cloud Gaming        & \phantom{1}8,430  & \phantom{1}80.0  & 2.93 & 77.07              & \phantom{1,}651.08  \\
        Industry 4.0        & \phantom{1}6,020  & \phantom{1}15.0  & 2.41 & 12.59              & \phantom{1,}708.60  \\
        Massive IoT         & \phantom{1}7,812  & \phantom{1}10.0  & 2.85 & \phantom{1}7.15    & \phantom{1,}839.35  \\
        Video streaming     & \phantom{1}9,241  & 100.0            & 3.09 & 96.91              & \phantom{1,}907.46  \\
        VoIP                & 13,458            & 100.0            & 0.70 & 99.30              & \phantom{1,}212.44  \\
        \hline
        \end{tabular}
        }
    \end{table}

    \begin{table}[ht]
    \centering
    \caption{Detailed Latency and Processing Statistics Per Service (Test Data)}
    \label{tab:latency_stats_test}
    \resizebox{\columnwidth}{!}{
    \begin{tabular}{|c|c|c|c|c|c|}
        \hline
        \multirow{2}{*}{\textbf{Services}} & \multirow{2}{*}{\textbf{Flows}} & \textbf{Mean Latency} & \textbf{Mean Proc.} & \textbf{Mean Margin} & \textbf{Mean Total} \\
         & & \textbf{Budget (ms)} & \textbf{Delay (ms)} & \textbf{(ms)} & \textbf{Bandwidth (Mbps)} \\
        \hline \hline
        Augmented reality   & \phantom{1}1,594  & \phantom{1}20.0   & 6.72 & 13.28 & 1,561.89 \\
        Cloud Gaming        & \phantom{1}6,009  & \phantom{1}80.0   & 3.79 & 76.21 & \phantom{1,}841.50  \\
        Industry 4.0        & \phantom{1}8,545  & \phantom{1}15.0   & 3.12 & 11.88 & \phantom{1,}917.04  \\
        Massive IoT         & \phantom{1}5,500  & \phantom{1}10.0   & 3.73 & \phantom{1}6.27  & 1,097.47  \\
        Video streaming     & \phantom{1}6,635  & 100.0             & 3.95 & 96.05 & 1,162.29  \\
        VoIP                & 18,778            & 100.0             & 0.91 & 99.09 & \phantom{1,}275.05  \\
        \hline
        \end{tabular}
        }
    \end{table}

    \noindent Each flow is defined by its service type, source/destination, required bandwidth, and maximum end-to-end latency constraint. More details on each service SFC can be found in Table \ref{tab:SFC_ziazet_2023} and in \cite{Ziazet_2023_FNFW}.
    
    \begin{table}[ht]
        
        \caption{SFC (values adapted from \cite{Ziazet_2023_FNFW})}
        \resizebox{\columnwidth}{!}{%
        \centering
        \begin{tabular}{|c|c|c|c|c|c|}
            \hline
            Service  &  Class  & SFC  & Bandwidth  & Latency  &  \# Request\\ 
                       
            \hline \hline
         Cloud  & \multirow{2}{*}{eMBB}    &  \multirow{2}{*}{UPF$_{\textsc{CG}}$- CU - DU - RU} & \multirow{2}{*}{4 Mbps} 
                & \multirow{2}{*}{80 ms} & \multirow{2}{*}{[40-55]} \\ 
             gaming (\textsc{CG}) & & & & & \\
             \hline
             Augmented  & \multirow{2}{*}{eMBB}    & \multirow{2}{*}{UPF$_{\textsc{AR}}$- CU - DU - RU} & \multirow{2}{*}{100 Mbps} 
                        & \multirow{2}{*}{20 ms} & \multirow{2}{*}{[1-4]} \\ 
             reality (\textsc{AR}) & & & & & \\
             \hline
             \multirow{2}{*}{VoIP}  & \multirow{2}{*}{eMBB}    & RU - DU - CU - UPF$_{\textsc{VoIP}}$ & \multirow{2}{*}{64 Kbps} 
                                    & \multirow{2}{*}{100 ms} & \multirow{2}{*}{[100-200]} \\ 
             & &  UPF$_{\textsc{VoIP}}$- CU - DU - RU & & &  \\
             \hline
             Video  & \multirow{2}{*}{eMBB}     & \multirow{2}{*}{UPF$_{\textsc{VS}}$- CU - DU - RU} 
                    & \multirow{2}{*}{4 Mbps}   & \multirow{2}{*}{100 ms}       & \multirow{2}{*}{[50-100]} \\ 
             streaming (\textsc{VS}) & & & & & \\
             \hline
             Massive    & \multirow{2}{*}{mMTC}    & \multirow{2}{*}{RU - DU - CU - UPF$_{\textsc{MIoT}}$} & \multirow{2}{*}{[1 -50] Mbps} 
                        & \multirow{2}{*}{10 ms} & \multirow{2}{*}{[10-15]} \\ 
             IoT (\textsc{MIoT}) & & & & & \\
             \hline
             Industry 4.0   & \multirow{2}{*}{uRLLC}    & RU - DU - CU - UPF$_{\textsc{I4.0}}$ 
                            & \multirow{2}{*}{70 Mbps} & \multirow{2}{*}{15 ms} & \multirow{2}{*}{[1-4]} \\
             (I4.0)         & & UPF$_{\textsc{I4.0}}$- CU - DU - RU &  &  & \\
             \hline
        \end{tabular}
        } 
        \label{tab:SFC_ziazet_2023}
    \end{table}


\section{Results}
\label{sec:results}
We now discuss the validation of the proposed {\NameProposal} models and algorithms. All the performance metrics presented are the average result across 5 experiments (for both {\RLBaseline} and {\NameProposal}). 

    \subsection{Metrics and Baseline}

    We evaluate two DRL agents: 
    (i) {\RLBaseline}, a single PPO agent using Maskable PPO \cite{Huang_2022} (which is an extension of PPO \cite{Schulman_2017} that handles dynamic action spaces) that sequentially routes all services; and 
    (ii) {\NameProposal}, our asynchronous multi-agent setup with one PPO agent per service interacting with a shared environment. 
    We report three families of metrics. 
    \emph{Grade of Service (GoS)} is the fraction of requests admitted and routed within budget and capacity constraints. 
    \emph{Latency} is end-to-end delay for granted flows (propagation plus processing at pre-placed VNFs). \emph{Wall-clock} captures training time to convergence and evaluation time on the same hardware. 
    Table \ref{tab:hp_gnn} summarizes the primary hyperparameters. 
    The multi-agent system uses smaller rollout fragments and minibatches per service to maintain responsive updates under asynchrony, and a slightly lower Generalizsed Advantage Estimator (GAE) $\lambda$ to reduce bias from stale advantages when actors advance at different paces. 
    This configuration mirrors the intended deployment: independent xApps acting and learning without global barriers.

    \begin{table}[ht]
        \centering
        \small
        \caption{Key hyperparameters of {\RLBaseline} and {\NameProposal} setup.}
        \begin{tabular}{|c|c|c|}
            \hline
            Param               & \RLBaseline & \NameProposal \\
            \hline\hline
            $\gamma$            & 0.99         & 0.99 \\
            $\lambda$ (GAE)     & 0.95         & 0.90 \\
            learning rate       & 3e$-$4       & 5e$-$5 \\
            policy epochs       & 10           & 5 \\
            rollout size        & 4,096        & 1,024 \\
            mini-batch size     & 128          & 64 \\
            \hline
            clip ($\epsilon$)   & \multicolumn{2}{c|}{0.2} \\
            entropy coeff.      & \multicolumn{2}{c|}{0.001} \\
            net (policy / value)& \multicolumn{2}{c|}{[256, 128] / [256, 128]} \\ 
            features dim (GCN)  & \multicolumn{2}{c|}{256} \\
            eval interval       & \multicolumn{2}{c|}{$\sim$50k steps} \\
            eval early stop     & \multicolumn{2}{c|}{over 5 evals w/ no improvement} \\
            \hline
        \end{tabular}
        \label{tab:hp_gnn}
    \end{table}

    Agents are evaluated every 50k steps by testing on a separate environment for 300 requests and recording the number of accepted requests, average reward, average latency, among other metrics. Training stops if the agent goes through 5 evaluations without improving its acceptance rate and average reward. For testing, a separate 24-hour traffic dataset with similar distribution patterns was used, ensuring evaluation under comparable but unseen traffic conditions (statistical differences are evident in Tables \ref{tab:latency_stats_training} and \ref{tab:latency_stats_test}).

    
    \subsection{Learning Dynamics and Wall-Clock}
    
\begin{figure}[ht]
        \centering
        \includegraphics[width=0.9\linewidth]{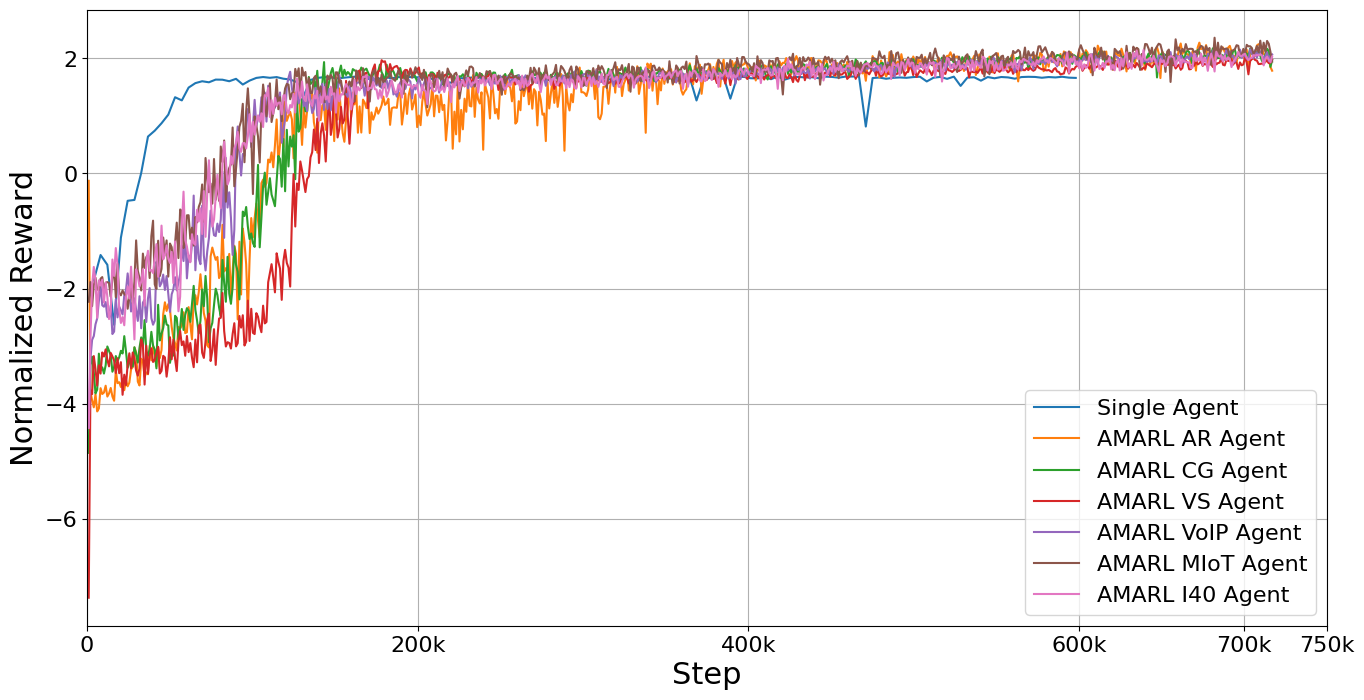}
        \caption{Learning curves for the {\NameProposal} and {\RLBaseline} during training.}
        \label{fig:training_reward_curves_single_AMARL}
        \Description{learning curves figure}
    \end{figure}

    Figure \ref{fig:training_reward_curves_single_AMARL} tracks learning progress. The {\RLBaseline} warms up quickly because all traffic funnels through one learner, but its exploration is serial and sensitive to stragglers. In contrast, the six service agents in {\NameProposal} start from heterogeneous operating points and move toward a common asymptote as the shared state stabilizes; their curves illustrate specialization early on and convergence later.

    The practical benefit of asynchrony appears in wall-clock time. Table \ref{tab:comparative_time_performance} shows that {\NameProposal} reduces training time from 07:25:05 to 05:12:00 (a 29.9\% reduction), and shortens test-time evaluation from 00:45:22 to 00:38:33 (a 15.0\% reduction). These gains come from parallel rollouts. 
    Let $\bar{L}$ be the average path length in hops and $\bar{b}$ the mean branching factor. A centralized single-agent explores
    $O(\bar{b}^{\bar{L}})$ choices per request; our design reduces wall-time by 
    \textit{(i)} partitioning traffic across $S$ agents (which means that each agent will see only its designed traffic resulting in a more homogeneous learning), and \textit{(ii)} running episodes and learning updates in parallel.

    \begin{table}[ht]
        \centering
        \caption{Comparative time performance}
        \begin{tabular}{|c|c|c|c|}
            \hline
           Methodology & Avg. GoS & Train & Test \\ 
            \hline
            \hline
            \RLBaseline  &  97.81\% & 07:25:05 & 00:45:22\\
            \NameProposal&  98.48\% & 05:12:00 & 00:38:33\\
            \hline
        \end{tabular}
        \label{tab:comparative_time_performance}
    \end{table}

    \subsection{GoS and Latency on Test Data}
\begin{figure}[!ht]
        \centering
        \includegraphics[width=0.85\linewidth]{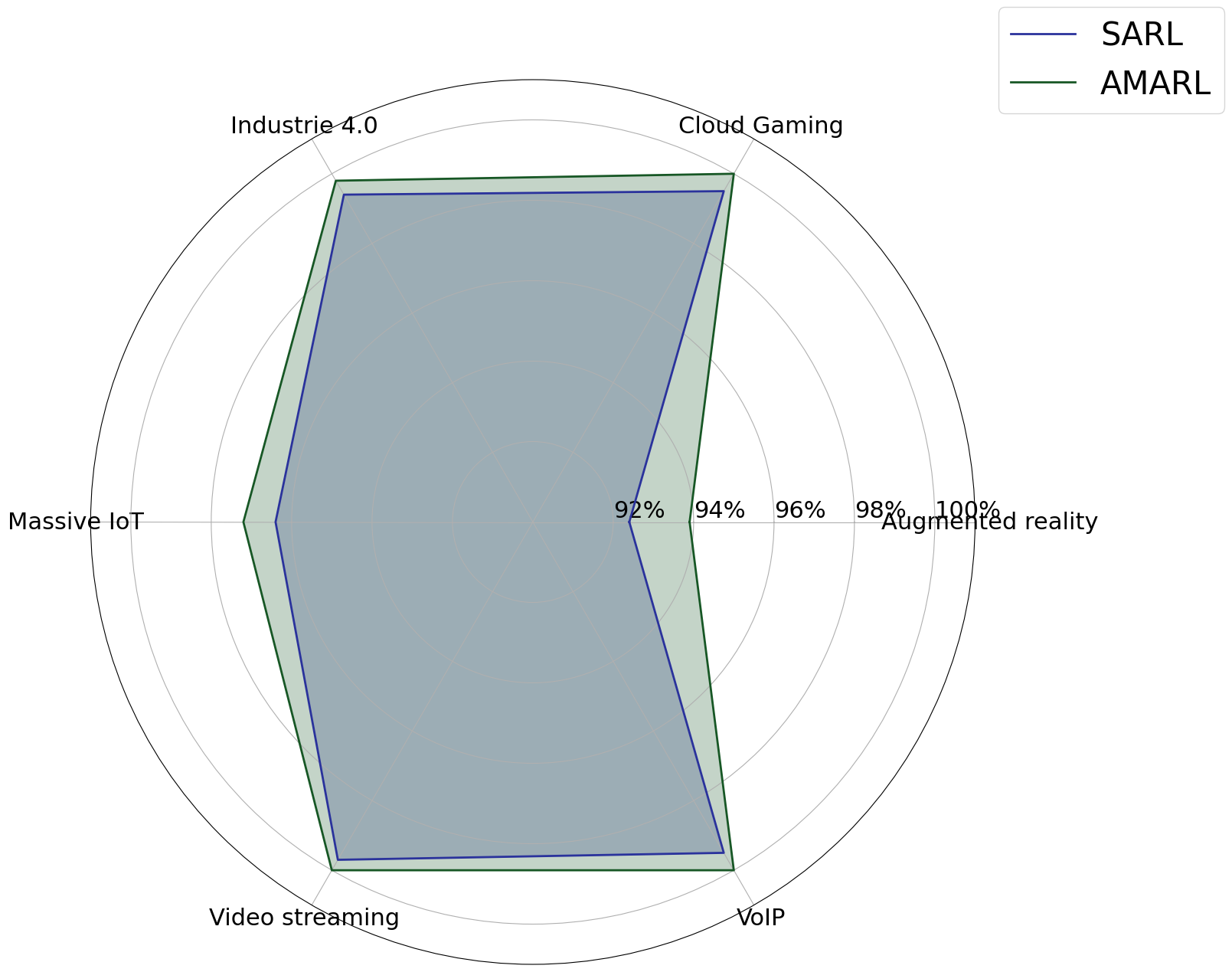}
        \caption{GoS comparison for both {\RLBaseline} and {\NameProposal} (Test Data)}
        \label{fig:AAMAS_GoS}
        \Description{GoS figure}
    \end{figure}

\begin{figure}[ht]
        \centering  \includegraphics[width=0.85\linewidth]{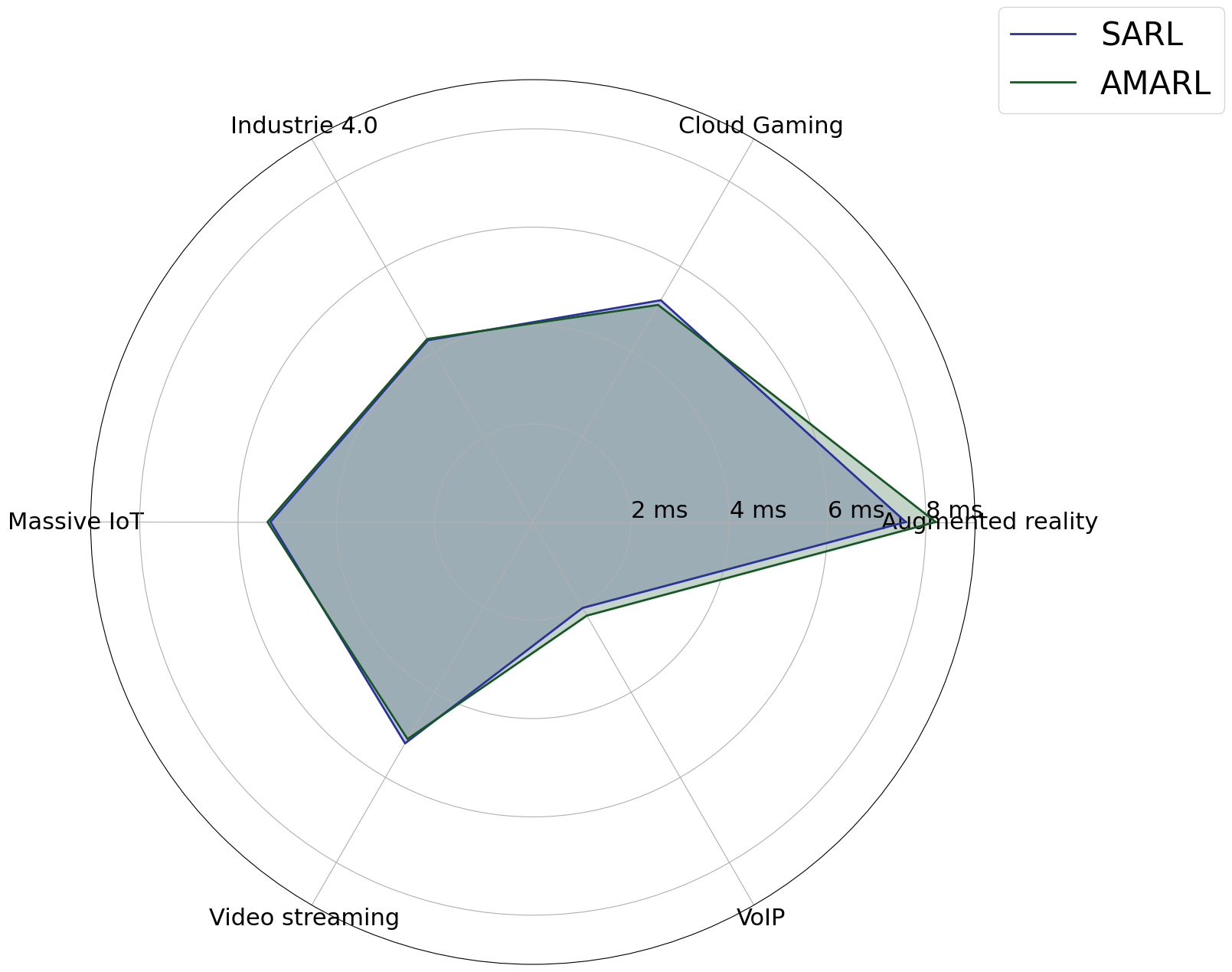}
        \caption{Latency comparison for both {\RLBaseline} and {\NameProposal} (Test Data)}
        \label{fig:AAMAS_Latency}
        \Description{Latency figure}
    \end{figure}

    At parity of resources, {\NameProposal} matches the single agent in terms of QoS. Figure \ref{fig:AAMAS_GoS} shows similar GoS at the aggregate level (\RLBaseline: $97.81\%$ vs.\ {\NameProposal}: $98.48\%$). Figure \ref{fig:AAMAS_Latency} compares service-level latencies and reveals that both methods deliver similar end-to-end delay envelopes across services. The AMARL points lie close to their single-agent counterparts, indicating that independence across service agents does not degrade routing quality under contention.
    
\begin{figure}[ht]
        \centering
        \includegraphics[width=0.7\linewidth]{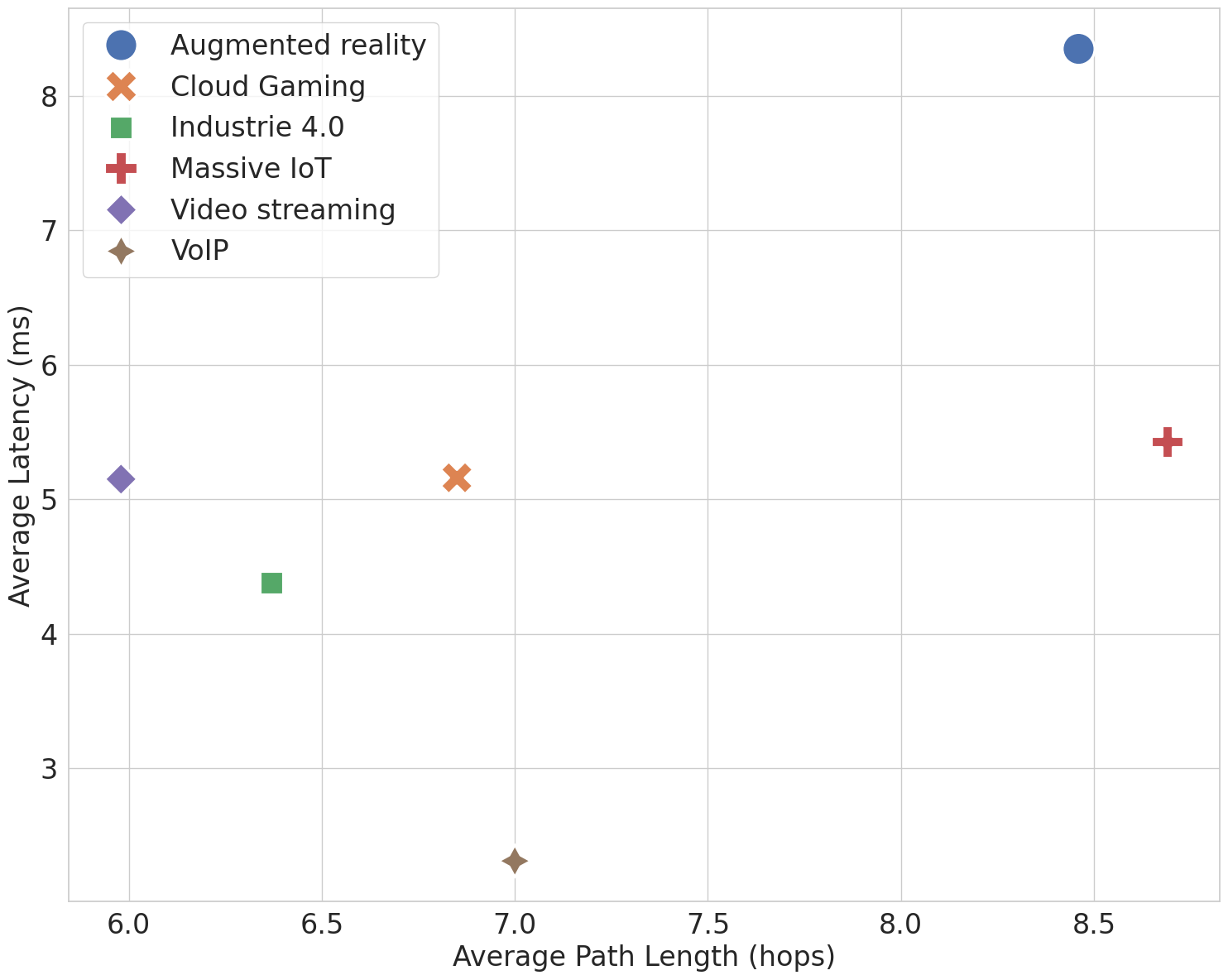}
        \caption{Latency Analysis Across Services (Test Data)}
        \label{fig:AAMAS_Latency_Analysis}
        \Description{Latency analysis figure}
    \end{figure}

\begin{figure}[ht]
        \centering
        \includegraphics[width=0.7\linewidth]{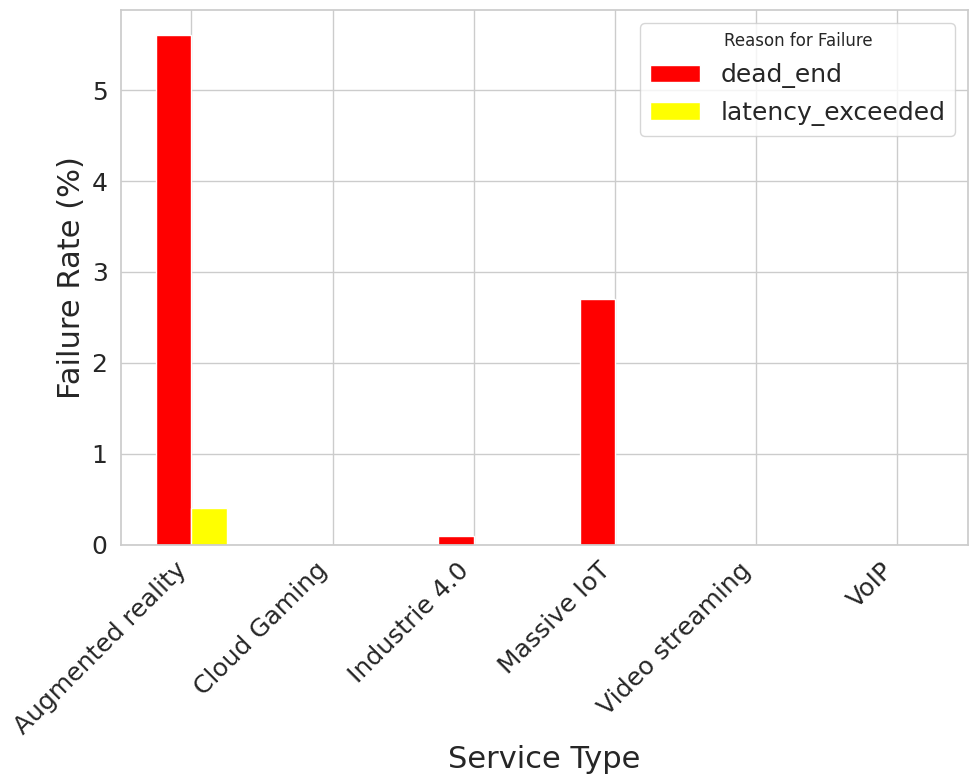}
        \caption{Failure Analysis (Test Data)}
        \label{fig:AAMAS_Failure_Analysis}
          \Description{Failure figure}
    \end{figure}

    Figure \ref{fig:AAMAS_Latency_Analysis} together with the test-flow statistics in Table \ref{tab:latency_stats_test} explains why some services exhibit longer average paths. Services with larger aggregate offered bandwidth (Mean Total Bandwidth) have a higher probability that the shortest path lacks sufficient residual capacity due to congestion; the action mask then prunes those hops, and the policy selects the following feasible detour. This capacity effect is visible when contrasting, for example, Video Streaming ($\sim$1,162\,Mbps) and Massive IoT ($\sim$1,097\,Mbps) against low-bandwidth VoIP ($\sim$275\,Mbps): the former tend to traverse longer paths on average, while the latter usually fits on the shortest path. Cloud Gaming (mid–high bandwidth, $\sim$842\,Mbps) shows intermediate behavior. Tight-latency services such as AR (budget 20\,ms; highest aggregate bandwidth $\sim$1,562\,Mbps) and I4.0 (15\,ms; $\sim$917\,Mbps) sit at the intersection of these forces: they often face capacity blocks on the shortest path, but their small latency margins (Mean Margin in Table \ref{tab:latency_stats_test}) limit how far they can detour. When no feasible hop remains within budget, episodes terminate as \texttt{dead\_end}, which is precisely the dominant failure mode observed in Figure \ref{fig:AAMAS_Failure_Analysis}. The higher failure rate for Augmented Reality (AR) and Massive IoT is a direct consequence of the masking semantics and their traffic profiles. A step is feasible only if at least one neighbor satisfies both residual-capacity and latency checks; otherwise the legal mask is empty and the episode ends as \texttt{dead\_end}. In the test set (Table~\ref{tab:latency_stats_test}), AR combines the highest offered bandwidth with a tight budget (20\,ms, small margin after processing), so shortest-path hops are frequently capacity–pruned; the remaining detours are then latency–pruned, leaving no legal move. Massive IoT has lower bandwidth but an even tighter budget (10\,ms), so it is predominantly latency–pruned even when capacity exists. By contrast, services such as VoIP pair low bandwidth with a generous 100\,ms budget, so shortest-path hops seldom violate capacity and feasible detours almost always remain. This bandwidth–budget interplay explains the pattern in Figure \ref{fig:AAMAS_Failure_Analysis}: AR is dominated by capacity\,+\,latency-induced \texttt{dead\_end}s, Massive IoT by latency-induced \texttt{dead\_end}s, while Cloud Gaming and Video Streaming fall in between.

    Overall, the parity of GoS/latency between {\NameProposal} and {\RLBaseline} indicates that distributing decisions across asynchronous service agents does not degrade routing quality under contention; it simply reallocates which flows take detours when shortest paths are congested, consistent with the bandwidth-driven feasibility constraints in the test set.


    \subsection{Results Discussion}

    The empirical results show that {\NameProposal} achieves parity with the single-agent baseline {\RLBaseline} on the two outcome metrics that matter most for operators---Grade of Service and end-to-end latency---while delivering substantially shorter training and evaluation wall-clock times (Table \ref{tab:comparative_time_performance}). This parity is non-trivial: in our setting, agents compete for shared resources under hard masks, and yet distributing decision-making across services does not degrade QoS. The data in Figures \ref{fig:AAMAS_GoS}--\ref{fig:AAMAS_Latency_Analysis} and Table \ref{tab:latency_stats_test} explain why: detours arise primarily from residual-capacity constraints on shortest paths, especially for high-bandwidth services, and both controllers learn to make similar trade-offs under contention.

    Beyond consistency, the asynchronous design offers several benefits that are challenging to achieve with a monolithic single agent, particularly in an O-RAN deployment use case like this one. First, asynchrony matches how real networks actually operate. The actions taken by operators are different depending on the service; a global clock does not synchronize them. {\NameProposal} mirrors this reality: each service agent reads the shared state, routes one request per local episode, and commits its diff independently. The aggregate effect is a steady flow of small decisions rather than bursts gated by barriers, which explains the $\sim$30\% reduction in training time and $\sim$15\% faster evaluation. In a live network, this may translate to quicker policy iteration.

    Second, the multi-agent layout improves fault tolerance and isolation. With a single global policy, learning or serving faults can cause the entire controller to stall. In our design, if one service agent backs off, crashes, or is rolled back, the others continue operating because coupling occurs only through the shared environment. The commit/abort mechanism confines failures to the affected service, and the shared-state guardrails ensure global feasibility despite local glitches. This aligns with operational practice where services are managed semi-independently to contain the blast radius.

    Third, {\NameProposal} yields specialization without interference. Each agent optimizes a reward tuned to its QoS envelope (e.g., stronger safety shaping for some services; softer for others) and adapts to its service’s demand patterns, as visible in the distinct learning trajectories of Figure \ref{fig:training_reward_curves_single_AMARL}. A single global agent must simultaneously internalize all objectives, which complicates credit assignment and renders policy updates brittle when the service mix shifts. In contrast, service-specific policies can be updated independently.

    Finally, the architecture is modular and extensible. As services in 5G---and soon 6G---evolve, introducing a new service does not require retraining a monolith. Operators can instantiate a new agent with a service-specific reward and observation head, train it in the same shared environment, and onboard it gradually (testing its behavior on a live network) while existing agents continue to serve traffic unchanged. Likewise, per-service A/B tests, rollbacks, or emergency patches do not destabilize other services. This modularity is a practical advantage for continuous integration of new xApps and for compliance workflows that demand service-level change management.

    
\section{Conclusion and Future Work}
\label{sec:conclusions_FW}

We presented {\NameProposal}, an asynchronous multi-agent reinforcement learning approach to service-specific routing in a shared network environment. Each service service runs an independent PPO policy that plans per-flow routes in a local episodic copy and commits resource deltas to a shared state, enforcing bandwidth, compute, and end-to-end latency constraints. Using an O-RAN–like testbed with fixed function placement, we showed that {\NameProposal} attains \emph{parity} with a strong single-agent baseline in Grade of Service and latency while delivering substantial wall-clock gains (about 30\% faster training and 15\% faster evaluation). Beyond metrics, the design aligns with near-real-time RIC practice (agents operate on different cadences), improves fault isolation, and enables specialization and modular evolution: new services can be onboarded as new agents without retraining a monolithic controller. In short,  we obtain these operational advantages while maintaining the performance (KPIs, satisfying QoS requirements) by adding minor complexity in network operations. We plan to (i) introduce priority-aware agents, so that more critical services preempt resources when contention arises, via priority-weighted rewards and commit arbitration policies; (ii) address non-stationary traffic through continual learning to mitigate catastrophic forgetting (e.g., replay windows, regularization-based consolidation, and drift detection for online adaptation); and (iii) explore lightweight inter-agent communication (explicit congestion and latency signals or learned message passing) to reduce avoidable detours while preserving the scalability of asynchronous operation.


\begin{acks}
This work was supported by NSERC (under project ALLRP 566589-21) and InnovÉÉ (INNOV-R program) through the partnership with Ericsson. We are grateful to Adel Larabi at GAIA, Ericsson Montréal for clarifying some concepts of the current 5G technology.
\end{acks}



\bibliographystyle{ACM-Reference-Format} 
\bibliography{Biblio/Intro, Biblio/literature, Biblio/BJ, Biblio/RL_libraries}


\end{document}